\documentclass[12pt,preprint]{aastex}

\slugcomment{ACCEPTED IN PASJ}
\begin{document}

\title{H$_{2}$ Formation in Low-Metallicity Galaxies}
\author{
Hideyuki K{\sc amaya}
\thanks
{Visiting Academics at Department of Physics, University of Oxford,
Keble Road, Oxford 1OX 3RH, UK.}
and Hiroyuki H{\sc irashita} \thanks
{Research Fellow of the Japan Society for the Promotion of Science.
 Visiting Researcher at Observatorio Astrofisico di Arcetri, Largo
 E. Fermi 5, Firenze, Italy.}
} 
\affil{Department of Astronomy, Faculty of Science, Kyoto University,
Sakyo-ku, Kyoto 606-8502} 
\email{kamaya@kusastro.kyoto-u.ac.jp}

\begin{abstract}
A possible formation mechanism of hydrogen molecules on a galactic scale
is examined. We are interested especially in the role of hydrogen 
molecules for the formation and evolution of primordial galaxies.
Thus, the formation process of hydrogen molecules in a very
low-metallicity galaxy (I Zw 18; the most typical
metal-deficient galaxy) is studied. 
Adopting a recent observational result of the absorption lines
of hydrogen molecules in I Zw 18, 
we obtain the upper limit for the ionization degree in the case
where hydrogen molecules can form via the H$^{-}$-process,
although they are generally believed to form on the surface of
dust grains. 
Furthermore, we present a critical ionization degree, 
above which the H$^{-}$-process can be dominant over
the formation process on the surface of grains. 
Interestingly, this critical ionization degree is comparable
to the upper limit of the ionization degree for I Zw 18.   
For determining the formation process of 
hydrogen molecules, future observational facilities can be useful.
Thus, we examine
the detectability in some wavelengths for metal-deficient
galaxies.
According to our estimate, the near-infrared line emission
of hydrogen
molecules is observable at the level of 10 $\mu$Jy,  
the free free radio emission is at the level of mJy,
and  the far-infrared emission from the dust on which hydrogen molecules
form  can also be detected at the 10-mJy level with its temperature of
16 K. The near-infrared line and the far-infrared continuum are
feasible for ASTRO-F observations.

\end{abstract}


\maketitle

\section{Introduction}

How efficiently hydrogen molecules, H$_2$,  form is the most
important topic for theories concerning the formation of
first-generation stars and primeval galaxies
(e.g., Matsuda et al. 1969). 
This is because
hydrogen molecules should become an important coolant in a
metal-deficient gas cloud.
A review article for such topics by the Japanese group is 
very useful for understanding the fundamental process in the 
formation of the first luminous objects via the formation of 
hydrogen molecules (Nishi et al.\ 1998).
As found in that review paper, there have been many theoretical research
projects for primordial hydrogen formation and the resultant effect.
The metallicity effect is also given in Omukai (2000). 
Observationally, 
there is little evidence for hydrogen formation
in primordial or very low-metallicity gas.
In this paper, we thus propose an idea to resolve this observational
weak point for a theory concerning 
the formation of primordial hydrogen molecules
on galactic and subgalactic scales. 

It is widely accepted that H$_2$ molecules are mainly formed
on dust grains in metal-rich galaxies 
(e.g., Williams 1993; Herbst 1995; Takahashi et al. 1999).
Low-metallicity galaxies, where only a little dust 
is expected to exist, are really important 
for our aim, since there is a
chance for the formation of molecules via the so-called
H$^-$-process (Lequeux, Viallefond 1980; Jenkins, Peimbert 1997). 
This process is the following: H$_2$ formation starts with 
the formation of a negative ion, H + e $\to$ H$^{-}$ + $h\nu$,
where H is hydrogen, e is an electron, H$^{-}$ is negatively and singly
charged hydrogen, $h$ is the Plank constant, and $\nu$ is the frequency
of the photon emitted during H$^-$ formation. 
That reaction rate is $1.0 \times 10^{-15}\, T_3 \exp (-T_3/7)$
cm$^3$ s$^{-1}$, where $ T_3$ is the temperature in units of $10^3$ K.
This process is followed by the {\it faster} associative
detachment reaction: H$^-$ + H $\to $ H$_2$ + e.
Hence, the formation rate of H$_2$ is mainly controlled by
H$^-$ formation. In our estimate, hence, we regard the formation rate 
of H$^-$ as the formation rate of H$_2$.
Throughout this paper, we call this formation process
of H$_2$ the H$^{-}$ process. 

While the reaction H$^-$ + H $\to $ H$_2$ + e occurs, 
the photodetachment of H$^-$ is also expected;
 H$^-$ + $h \nu $ $\to $ H + e.
Indeed, the H$_2$ formation rate via the above process is
$2\times 10^{-9} n({\rm H}) $ s$^{-1}$, while that detachment
occurs with a rate of $1.7 \times 10^{-7}$ s$^{-1}$
(Mathis et al.\ 1983). 
Thus, if the density of the interstellar medium is not very large
($< 100$ cm$^{-3}$),
it is very difficult for H$_2$ to form via the H$^{-}$-process.
Furthermore, by mutual neutralization, the formation of H$_2$
can also  be suppressed. 
That reaction is described as 
H$^-$ + X$^+ $ $\to $ H$_2$ + X,  
where X  is an arbitrary atom including hydrogen,
and its reaction rate is 
1.3$\times 10^{-7} T_3^{-0.5} X_{\rm e} n({\rm H})$ s$^{-1}$, 
where $X_{\rm e}$ is the ionization degree of the interstellar medium (ISM),
$n({\rm H})$ is the number density of neutral hydrogen,
and $T_3$ is temperature in units of $10^3$ K.
In the subsequent discussion, we can regard $X_{\rm e}$
as being much smaller than unity, and thus the latter process is slower
than the photodetachment.
Hence, we consider only the photodetachment process 
against the formation of a hydrogen molecule via the H$^{-}$-process. 

In this paper, 
we discuss whether the H$^{-}$-process for H$_2$ formation 
occurs in an observable low-metallicity galaxy. 
Our consideration will be useful for the theory of galaxy formation
and first-generation stars when a metal-deficient galaxy is 
found to be a laboratory for  H$_2$ formation via the
H$^{-}$-process.
In section 2, we review an interesting observation of I Zw 18,
which is the lowest known metallicity galaxy showing present
star-forming activity.
In section 3, we present an upper limit for the ionization degree,
adopting the observational constraint of I Zw 18.
If that upper limit is reached,
the H$^{-}$-process is expected to be the dominant process of
H$_2$ formation in such a low-metallicity galaxy.  
In section 4, we give implications for future observations.
Finally, we summarize this paper in section 5.

\section{Observation of a Metal-Deficient Galaxy}

Metal-deficient galaxies are suitable for our aim to find
evidence of H$_2$ formation via the H$^{-}$-process on  
galactic and subgalactic scales. 
The most typical and famous metal-deficient galaxy is I Zw 18.
This is classified as a blue-compact dwarf (van Zee et al.\ 1998) 
and shows evident star formation. Importantly, I Zw 18
has the smallest abundance of heavy elements found in the
ionized region in galaxies. Moreover, CO is not detected (e.g.,
Gondhalekar et al.\ 1998) in it. Although the latter does not always 
mean underabundance of C and O, it is just consistent with
the metal-deficient property of I Zw 18. Hence, we regard this galaxy
as being a typical metal-deficient galaxy.
We note that I Zw 18 has another name, Mrk 116.

Recently,  Vidal-Madjar et al.\ (2000) observed  I Zw 18 
by using the {\it Far Ultraviolet Spectroscopic Explorer}. 
Importantly, they did not succeed
in detecting absorption feature of hydrogen molecules. Thus, the upper
limit of the column density of hydrogen molecules is
determined. That is, $N({\rm H}_2) < 10^{15}$ cm$^{-2}$, 
where $N({\rm H}_2)$ is the column density of hydrogen molecules.
Assuming the distance to I Zw 18 to be 11.5 Mpc, 
we adopt a size of I Zw 18, $R_{18} = 1.7$ kpc, which is proposed 
in van Zee et al.\ (1998).
We can thus determine the upper limit of the number density of 
hydrogen molecules as being 
$n({\rm H}_2) < 1.9 \times 10^{-6}$ cm$^{-3}$.
In this paper, using this constraint for $N({\rm H}_2)$ 
and $n({\rm H}_2)$,
we consider the formation mechanism of hydrogen molecules in a
metal-deficient galaxy. 

For our consideration in this paper, it is very important
for us to estimate the strength of the radiation field of I Zw 18.
According to figure 2 of Dufour et al. (1988),
we find the observational flux at 1300 \AA  ~as being
$\sim 2 \times 10^{-14}$ erg  s$^{-1}$ cm$^{-2}$  \AA$^{-1}$,
which was determined by means of        
the {\it International Ultraviolet Explorer} satellite. 
The distance to I Zw 18 is assumed to be $\sim 10$ Mpc 
and the size of of star-forming region of I Zw 18 to be $\sim 1$ kpc.
Both quantities were applied to find a rough estimate of 
the mean radiation field of I Zw 18. 
Adopting them, we found that  
the mean radiation field at the same wavelength
is about $\sim 2 \times 10^{-8}$ erg  s$^{-1}$ cm$^{-2}$ \AA$^{-1}$,
which is consistent to the radiation 
field adopted in Vidal-Madjar et al.\ (2000). 
In a recent work by Stasi\'nska and Schaerer (1999), 
it has been confirmed. In addition, Stasi\'nska and Schaerer
have suggested that ISM in I Zw 18 
can be clumpy, although the dense clumps have only a small volume filling
factor. In the next section, we also consider a clumpy ISM whose clumps
have a density of 100 cm$^{-3}$ as suggested in Stasi\'nska and Schaerer. 

We can put a constraint on the star-formation history of I Zw 18
from the elemental abundance and ionization, both of which 
were measured in an H {\sc ii} region. 
Importantly, the expansion
of super-bubbles can also affect chemical evolution, because
newly synthesized elements are expelled to inter-galactic space
(Martin 1996). 
This consideration has been developed further to other
metal-deficient galaxies (Martin 1997).
The effect of the out-flow will be examined in a forthcoming
paper. In the current paper, we try to connect the clumpiness of ISM
and the intense radiation field, to reveal the formation processes
of molecular hydrogens under the constraint given in Vidal-Madjar et
al.\ (2000).

\section{Upper Limit for the Ionization Degree}

\subsection{Assumptions}

We consider the two branches for the formation of hydrogen molecules.
As introduced in section 1, the first one is
the H$^{-}$-process. The second one is expected on the surface of
dust grains (Hollenbach et al.\ 1971). We  call the latter case
the dust-process in this paper. 
Since we are interested in the H$^{-}$-process,
we examine the possibility of the H$^{-}$-process in I Zw 18, which
is a typical metal-deficient galaxy.

Here, we present our assumptions.
Two important assumptions are employed:  
(i) The interstellar medium (ISM) is clumpy.
The size of a typical clump is about 1 pc and its gas-number density
is about 100 cm$^{-3}$.
It is observationally believable for 
such a very small-scale structure to exist
(e.g., Frail et al.\ 1994).   
To sum up, 
we are interested in a clumpy H {\sc i} medium whose
{\it mean} number density may be about 0.4 cm$^{-3}$
(Vidal-Madjar et al.\ 2000).   
(ii) A steady state between the formation and destruction of hydrogen
molecules is
established. Hence, we estimate the number fraction of hydrogen
molecules from an equilibrium condition.
The number fraction of hydrogen molecules, $f({\rm H_2})$, is
defined to be the ratio of the number of hydrogen molecules to
the total number of hydrogen nuclei.
Although it may not be so significantly important, we state a third 
assumption: 
(iii) $f({\rm H_2})$ is approximately  $n({\rm H_2})/n({\rm H})$. 
This is very reasonable if most of the hydrogens are in the
form of neutral hydrogen.

The first assumption makes it possible that the photodetachment process  
is suppressed. If we considered the condition of only the diffuse ISM,
it would be found that the H$^{-}$-process is not efficient owing to 
the photodetachment destruction of H$^{-}$. However, in the assumed
clumps, the formation process of H$_2$ via the H$^{-}$-process can 
occur faster than the photodetachment.
If we adopt the Galactic condition, in each clump, whose size
is about 1 pc, its $A_V$ is estimated to be about 4. 
In such a condition,
a standard PDR (photodissociation region) model 
predicts an ionization degree of $\sim 10^{-4}$
(Hollenbach, Tielens 1999).
Someone might think that the Galactic values are inadequate for
I Zw 18 owing to its metal deficiency.
Fortunately, however, the lower metallicity condition 
for a fixed $A_V$ is suitable
for the larger size of the assumed clumps.
When clumps with a fixed size of 1 pc 
exist in the low-metallicity condition, it is required that
they have a large ionization degree.
Then, our first assumption will be considered as a standard case
when we consider whether the H$^{-}$-process works or not.

We also comment on the volume fraction of the small clumps, 
$p_{\rm fr}$. If we assume that the interstellar gas is composed of clumpy
and diffuse components, and that the density of the diffuse component
is 0.1 cm$^{-3}$, we estimate $p_{\rm fr} \sim 0.003$ from the instant 
estimate of $0.1 \times (1- p_{\rm fr}) + 100 \times p_{\rm fr} =0.4$.
Thus, if the mean density of the H {\sc i} medium is 0.4 cm$^{-3}$, as
given in Vidal-Madjar et al., the clump fraction is very
small. Although our theoretical consideration is not altered as long
as $p_{\rm fr}$ is much smaller than unity, the observational 
implications presented in section 4 are affected.  
Fortunately, I Zw 18 has a H {\sc i} envelope, which is
characteristic of blue compact dwarfs.  In the H {\sc i} envelope,
$p_{\rm fr}$ could be on the order of 0.1. Then, 
the formation of H$_2$ would be observable in the H {\sc i} 
envelopes of the blue compact dwarfs with metal deficiency. 

The second assumption holds if the time-scale of 
gas consumption into stars is sufficient long.
That is, the changing time-scale of the amount of heavy metal 
and dust is assumed to be longer than the interesting
time-scale needed to establish equilibrium between the destruction
and formation of molecular hydrogens.
Since we are interested in the evolution of present-day galaxies,
the time-scale of the variation of heavy metals and dust 
is approximately equal to the duration of star formation
on a galactic scale. The variation time-scale for the star-formation rate 
may be about 10$^9$ yr (e.g., Legrand et al.\ 2000). Thus, the
destruction time-scale
and the formation time-scale of molecular hydrogens should be shorter
than 10$^9$ yr. We confirm this point in the last paragraph 
of this section, since we need a quantitative definition for the formation 
and destruction of hydrogen molecules.

\subsection{H$^{-}$-Process}

Here, we pay attention to the formation process
via the H$^{-}$-process.
First of all, we examine the physical condition of clumps
in which H$_2$ forms via the H$^{-}$-process. The
density and temperature are assumed to be 100 cm$^{-3}$ and
100 K, respectively. 
Those parameters are taken from the observational indication 
of the cold atomic gas in the Galaxy
(see Kulkarni, Heiles 1988; Dickey, Lockman 1990 for reviews).
As long as the realistic condition of H {\sc i} medium
in I Zw 18 is in unclear, the adopted condition for the clumps
is just the assumptions.
By the way, that density is very interesting. 
This is because we can safely
expect that H$_2$-formation via the H$^{-}$-process 
against photodetachment occurs above a density of 100 cm$^{-3}$.
That is, that density is also estimated from the reaction rates 
of the both processes presented in the third paragraph of section 1.
Hence, we adopt a typical density of the clumps as
being 100 cm$^{-3}$ for a critical density. 

Defining $R_{\rm H^{-}}$ to
be the formation rate of hydrogen molecules in the dimension of
cm$^3$ s$^{-1}$, we can estimate the formation rate of
hydrogen molecules to be $2R_{\rm H^{-}} n({\rm H}) n_{\rm e}$, where
$n_{\rm e}$ is the number density of electrons. 
For the convenience of subsequent sections,
we rewrite the formation rate in terms
of the ionization degree, $X_{\rm e} = n_{\rm e}/n({\rm H})$; 
it then becomes
$2R_{\rm H^{-}} n({\rm H})^2 X_{\rm e}$. In terms of the ionization degree,
we can discuss a general condition of interstellar clouds, as found
in the next section.

The destruction of hydrogen molecules is considered to 
occur efficiently owing to photodissociation.
We thus estimate the destruction rate in the dimension of
cm$^{-3}$ s$^{-1}$ as $I \times n({\rm H_2})$, where $I$ is  
the photodissociation rate. As a result, we adopt the following
equilibrium condition for the above quantities:
\begin{eqnarray}
I \times n({\rm H_2}) = 2R_{\rm H^{-}} n({\rm H})^2 X_{\rm e},
\label{eq:hminus}
\end{eqnarray}
that is,
\begin{eqnarray}
f({\rm H_2}) = \frac{2R_{\rm H^{-}} n({\rm H}) X_{\rm e}}{I}.  
\end{eqnarray}

As a destruction mechanism of molecular hydrogen, we assume
photodissociation owing to  massive stars. Since I Zw 18
shows evident star-forming activity, our assumption is reasonable.
To make our presentation clear, moreover, we adopt the same radiation
field of Vidal-Madjar et al.\ (2000). That is, $I = 4 \times 10^{-11}$
s$^{-1}$ is employed. Importantly, the assumed clumps have
an extinction of about $A_V \sim 4$. Then, the adopted $I$ is rather
overestimated for the destruction of H$_2$.
In other words, the determined $f({\rm H_2})$ is underestimated.
Hence, even if our $f({\rm H_2})$ is comparable to the limit
for the H$^-$-process to be active,
we still expect that H$_2$ forms via the H$^-$-process.
Finally, we present only the conclusions with this constraint.

We should note that
this photodissociation rate is expected near the O9.5 V star, i.e.,
$\zeta$ Oph (Jura 1974). This corresponds to the strength of the radiation
field with an intensity of
$G = 3 \times 10^{-8}$ photons cm$^{-2}$ s$^{-1}$ Hz$^{-1}$
(i.e. $2\times 10^{-6}$ erg cm$^{-2}$ s$^{-1}$ {\AA}$^{-1}$) at
1000 \AA.
We want to insist that since such an intense radiation field exists,
hydrogens are partially ionized in the H {\sc i} medium 
(e.g., Reynolds et al.\ 1998). 
If electrons originate from
hydrogens, the electron number density is not so significantly
small as that in the case of ionization of only the heavy elements
(e.g., carbon). Therefore, if we considered the origin of electrons
as only carbon,
we would underestimate the number density of electrons.
The formation rate of hydrogen molecules is estimated to be 
\begin{eqnarray}
R_{{\rm H}^{-}} 
= 1.0 \times 10^{-15}T_3 \exp ( -{T_3}/{7} )
~~{\rm cm^3 ~ s^{-1}},
\end{eqnarray}
where $T_3$ is the temperature in units of $10^3$ K.
For 100-K clumps,
$R_{{\rm H}^{-}}=1.0\times 10^{-16}~{\rm cm^3 ~ s^{-1}}$.

{}From the above estimate, 
we can determine the fraction of molecular hydrogen, $f({\rm H}_2)$,
assuming the temperature of the gas clumps to be 100 K. 
Adopting the upper limit of the column density of hydrogen
molecules and the typical column density of neutral hydrogen
[$N({\rm H}) = 2\times 10^{21}$ cm$^{-2}$;
Vidal-Madjar et al.\ 2000], $f({\rm H}_2)<10^{-6}$ is obtained.
We thus find $X_{\rm e}$  to satisfy
the following condition via equation (2):
\begin{eqnarray}
   X_{\rm e} < 2 \times 10^{-3} 
\times \left[ \frac{10^{2} {\rm cm}^{-3}}{n({\rm H})} \right] .
\label{eq:upper1} 
\end{eqnarray}
Since the upper limit of the left-hand side of equation (2) is
provided by Vidal-Madjar et al.\ (2000) and section 2, 
we obtain the resultant upper
limit for $X_{\rm e}$ for a given
$n({\rm H})$, which is estimated to be 100 cm$^{-3}$ in inequality
(\ref{eq:upper1}). 
This upper limit is allowed as long as the clumps are not very large.

We stress again that such partial ionization
of hydrogens is possible,  
since there are ionizing stars in I Zw 18.
By the way, we still do not judge whether the H$^{-}$-process works well
in such a metal-deficient galaxy.
Even if the ionization degree of I Zw 18 
is estimated to be about the upper limit of
the right-hand side of inequality (4),
the dust process may be dominant over the H$^{-}$ process. 
Thus, we should examine whether the H$^{-}$ process really
dominates over the dust process. 
We consider this topic in the next subsection.

\subsection{Dust-Process}

We know the formation rate of  molecular hydrogen via the dust-process,
\begin{eqnarray}
R_{\rm dust} = 10^{-17} ~~ {\rm cm^{3}~ s^{-1}}. 
\end{eqnarray}
This is the mean value expected in our Galaxy and found by Jura (1974)
to be an upper limit. The equilibrium state becomes
\begin{eqnarray}
I \times n({\rm H_2}) = 
2R_{\rm dust} n({\rm H})^2  \frac{\cal D}{6\times 10^{-3}}, \label{eq:dust}
\end{eqnarray}
that is,
\begin{eqnarray}
f_{\rm dust}({\rm H_2}) = \frac{2R_{\rm dust} n({\rm H})}{I}
		  \frac{\cal D}{6\times 10^{-3}}.  
\end{eqnarray}
Here, ${\cal D}$ is the dust gas ratio in mass and the displayed quantity
of ${\cal D}$ in equations (6) and (7) is the Galactic value
(Spitzer 1978).

We wish to note that 
$f_{\rm dust}({\rm H_2})$ depends on the dust-gas ratio.
That is, if there were no dust in interstellar space, 
$f_{\rm dust}({\rm H_2})$ 
would always be zero. 
Considering the dust-process, 
Vidal-Madjar et al.\ (2000) have
estimated another upper limit of the column density of hydrogen
molecules that satisfies their own observational constraint.
In this meaning, their discussion is consistent.
In this paper, we undertake another
discussion by using the same observational constraint. 

Here, we examine the condition that the H$^{-}$-process is dominant
over the dust process. 
Equating the left-hand sides of equations (\ref{eq:hminus}) and 
(\ref{eq:dust}),
we  determine a critical ionization degree of $ X_{\rm ec}$
with values of $R_{{\rm H}^-}$ and
$R_{\rm dust}$, 
\begin{eqnarray}
X_{\rm ec} = 1\times 10^{-1} 
	    \times \left(\frac{\cal D}{6\times 10^{-3}}\right).  
\end{eqnarray}
If $ X_{\rm e} > X_{\rm ec}$, 
the H$^{-}$-process dominates over the dust-process for a given radiation
field. If not,
the dust-process is the dominant formation process of hydrogen
molecules. Thus, we present the first conclusion as follows. Assuming
${\cal D}$ to be about $10^{-4}$ in I Zw 18 (about 1/50 of the
Galactic value), $ X_{\rm ec}$ is estimated
to be $2 \times 10^{-3}$. We have already obtained the upper limit
of $X_{\rm e} $ of I Zw 18 in equation (\ref{eq:upper1}) as
$2 \times 10^{-3}$ for $n({\rm H}) = 100$ cm$^{-3}$.
This upper limit is comparable to  $ X_{\rm ec}$. 
This means that the H$^{-}$-process is expected to occur in the
small H {\sc i} clumps in I Zw 18 if the upper limit of $X_{\rm e} $ is
reached.

Before closing this section, we examine whether an equilibrium state
is established
well on a typical time-scale of galaxy evolution. The destruction
time-scale of H$_2$ is about $I^{-1}
\sim 10^3$ yr. As long as equations (\ref{eq:hminus}) and
(\ref{eq:dust}) are adopted, the formation 
time-scale should also be of the same order of magnitude due to the
definition of equilibrium.
Fortunately, the evolution time-scale (i.e., time-scale of metal
enrichment) of star-forming galaxies is $> 10^9$ yr (e.g.,
Legrand et al.\ 2000). 
Thus, our steady state assumption is reasonable for our research.

\section{Observational Implication}

\subsection{H$_2$ Emission Line in the Near Infrared}

Observations of the H$_2$ rotational and vibrational line emissions  
are important
to show the formation of molecules in clumps, as assumed in
section 3.
Here, we adopt the model of photodissociation regions by
Black and van Dishoeck (1987). To be consistent with the
density and temperature adopted in section 3, we adopt Model 1, 
whose physical parameters are listed in their table 1.
[It is assumed that $n(H)=1.0\times 10^2$ cm$^{-3}$ and $T=100$ K.
For the ultraviolet flux, $G\sim 2\times 10^{-6}$ erg cm$^{-2}$
s$^{-1}$ {\AA}$^{-1}$, which is comparable to that at
$\zeta$ Oph, is adopted (section 3.2).] From their tables 1 and
2, we see that the intensity of the line at 2.406 $\mu$m is
$I_{2.4}=9.0\times 10^{-8}$ erg s$^{-1}$ cm$^{-2}$ sr$^{-1}$. We
have chosen this line because it is the strongest.

We estimate the flux detected on the earth, $f_{2.4}$
(erg s$^{-1}$ cm$^{-2}$ Hz$^{-1}$). This is estimated  to be
\begin{eqnarray}
f_{2.4}=\frac{I_{2.4}\Omega}{\Delta\nu},
\end{eqnarray}
where $\Omega$ and $\Delta\nu$ are the solid angle of the object
and the typical width of the wavelength band. 
For a particular interest in metal-deficient galaxies, we
estimate the flux for I Zw 18, whose
distance is assumed to be 11.5 Mpc and the typical radius for the
gas envelope is 1 kpc (i.e., $\Omega =2.5\times 10^{-9}$).

To determine $\Delta\nu$, we specify the observational facility
if the wavelength resolution is larger than the typical
broadening of the line.
Here, we estimate the detectability by ASTRO-F. ASTRO-F
is planned to conduct deep imaging in the near-to-mid infrared
range by the infrared camera (IRC;
http://www.ir.isas.ac.jp/ASTRO-F/index.html; Shibai 2000;
Onaka 2000).\footnote{ASTRO-F
will also make an all-sky survey in the far infrared.}
The wavelength resolution at near infrared (NIR) is
$\nu/\Delta\nu\sim 40$. Thus, at 2.4 $\mu$m,
$\Delta\nu\sim 3\times 10^{12}$ Hz. Since this is much larger
than the typical broadening ($\sim 4\times 10^9$ Hz for the velocity
dispersion of 10 km s$^{-1}$), the line width is determined by
the resolving power. Adopting $\Delta\nu\sim 3\times 10^{12}$ Hz,
we obtain $f_{2.4}\sim 8~\mu$Jy
(1 ${\rm Jy}= 10^{-23}$ erg s$^{-1}$ cm$^{-2}$ Hz$^{-1}$). Since the
detection limit of
IRC is $\sim 10~\mu$Jy, it may be marginally possible to detect
the NIR line of H$_2$ molecules.
The molecules may suffer stronger ultraviolet (UV) radiation
if they exist near star-forming regions.
If Model 2 of Black and van Dishoeck (1987) is adopted,
where the UV intensity is 30-times larger than that in Model 1,
we obtain $f_{2.4}\sim 50~\mu$Jy
($I_{2.4}=5.7\times 10^{-7}$ erg s$^{-1}$ cm$^{-2}$ sr$^{-1}$). In
this case, the line will be 
easily detected by future observations. However, we should note
that if the covering factor of the H$_2$ molecule is much smaller than
unity, the detection becomes difficult.
In summary, the detection of the H$_2$ molecular line from 
I Zw 18 by ASTRO-F
indicates that the covering factor of the molecule is 
not much smaller than unity
and that the UV radiation field is stronger than the value
for $\zeta$ Oph. As already commented in subsection 3.1, 
we predict that 
such H$_2$ emission is observed in the H {\sc i} envelopes
of metal-deficient blue compact dwarf galaxies 
around the periphery of the star-forming region and/or at the inner edge
of the H {\sc i} envelope. 

However, if the line-to-continuum ratio might be so distinct,
greater spectroscopic resolution will be needed to find
the line emission of molecular hydrogen in I Zw 18. 
In this meaning, the IRCS instrument on board the SUBARU telescope
may be useful, while if the expected surface brightness
is very small ($\sim $ 0.5 $\mu$Jy arcsec$^{-2}$),
the current observational facility will have  difficulty 
to observe it. If it is observed, the hydrogen formation
will be enhanced by the dust. This will constrain 
the dust evolution model of dwarfs (Hirashita 1999).


\subsection{Radio Continuum}

If the ionization degree is large (e.g., $X_{\rm e} > X_{\rm ec}$), we may
expect the radio free free
emission from H {\sc i} envelopes of the metal deficient galaxies. 
Someone might think that the contamination of electrons in
the star-forming region of I Zw 18 exists. We admit that difficulty.
Fortunately, however, 
I Zw 18 is a blue compact dwarf galaxy, which is often associated
with an extended H {\sc i} envelope. 
Indeed, such an envelope is observed around I Zw 18
(Lequeux, Viallefond 1980).
Thus, we pay attention to the free free radiation 
from the H {\sc i} envelope of I Zw 18. Of course, it is only
partially ionized owing to its central radiation field 
and stars in the H {\sc i} envelope, itself, 
although we do not know the exact ionization 
degree of the H {\sc i} envelope at this moment.

The emissivity of the free free emission is 
$$
\epsilon_\nu^{\rm ff} = 
6.8 \times 10^{-38}~~~ {\rm erg~cm^{-3}~s^{-1}~Hz^{-1}} $$
\begin{eqnarray}
\times n_{\rm e} n_{\rm i} T^{-0.5} 
{\rm exp}\left(- \frac{h\nu}{k_{\rm B}T}\right), 
\end{eqnarray}
where $n_{\rm i}$ is the ion number density, 
and $k_{\rm B}$ is the Boltzmann
constant (Rybicki, Lightman 1979). The Gaunt factor is estimated to
be of order unity. First, we treat a diffuse media.
Since the ionization of hydrogen is assumed to be dominant
over the other species in the supply of electrons, $n_{\rm i}$ is estimated
to be about $n_{\rm e}$.
We adopt $n_{\rm e}$ as being about
$n({\rm H})X_{\rm ec}=8\times 10^{-4}$ cm$^{-3}$ for
$n({\rm H}) \sim 0.4$ cm$^{-3}$,
we find it for a radio wavelength ($h\nu\ll k_{\rm B}T$),
\begin{eqnarray}
 F_\nu^{\rm ff} \sim 1\times 10^{-10}~{\rm Jy},
\label{eq:radioflux}  
\end{eqnarray}
where $F_\nu^{\rm ff}$ is the flux of free free emission in the
electron temperature of 7000 K. We assume the size of the H {\sc i}
envelope to be 2 kpc (i.e., $>R_{18}$) and 
the distance to I Zw 18 to be 11.5 Mpc. The flux level is impossible
to detect it, since
at most mJy flux will be observable by means of  observational facilities
in the near future.

The above estimate has been applied to a diffuse medium.
Next, we consider the clumpy H {\sc i} medium in a consistent way 
with the previous sections. 
Interestingly, such dense clouds can form in the H {\sc i}
envelope (e.g., Sait\={o} et al.\ 2000). That is, there may be a
significant amount of 
dense clouds in the H {\sc i} envelope. To make the situation simple,
almost all hydrogens are hypothesized to become dense clouds 
in the H {\sc i} envelope. In such a case, we expect the free free
radiation in a radio band to be on the order of 10 mJy, adopting
$n({\rm H})=10^2$ cm$^{-3}$,
$X_{\rm e} \sim 2\times 10^{-3}$, and $T=100$ K. 
Thus, we suggest that
if intense free-free emission is observed in the {H{\sc i}} envelope,
we can expect that hydrogen molecules form in clumps in that envelope via 
the H$^{-}$ process. To justify this, of course, we should determine 
the amount of dust in the H {\sc i} envelope, 
since it is necessary for $\cal D$ to be very small. We then discuss
the observational feasibility in the far-infrared (FIR) band, where
the thermal emission of dust has a peak intensity, in order
to constrain the effectiveness of the H$^{-}$ process
in metal-deficient galaxies.

\subsection{Far-Infrared Band}

Here, we examine the observational feasibility in the FIR band.
%
To be consistent with the above sections, we set
the dust-to-gas ratio in I Zw 18 as
${\cal D}=10^{-4}$. If the FIR radiation is not detected
with the level of the result later, the dust-to-gas ratio
should be less than $10^{-4}$. In such a case, $X_{\rm ec}$
becomes lower, and the possibility that the H$^{-}$-process
is dominated over the dust-process becomes higher.
Other parameters are selected as follows:
the {\it mean} number density of hydrogen is $0.4$ cm$^{-3}$, 
and the typical radius of the H {\sc i} envelope is 1 kpc.
We note that both of the parameters are consistent with the above
discussions.

Adopting the above parameters, we estimate the total mass of dust
in I Zw 18 to be $M_{\rm dust}\sim 8\times 10^3 M_\odot$.
Lonsdale Persson and Helou (1987) related $M_{\rm dust}$
and $f_\nu$ (the flux of the dust emission at the frequency of
$\nu$) as
\begin{eqnarray}
f_\nu =\frac{K_\nu M_{\rm dust}B_\nu (T)}{D^2},
\end{eqnarray}
where $K_\nu$ is the emissivity of dust per mass,
$B_\nu (T)$ is the Planck function, and $D$ is the distance
to I Zw 18.
They also estimated $K_\nu$ at a wavelength of 100 $\mu$m as
90 cm$^2$ g$^{-1}$ based on Draine and Lee (1984).
The dust temperature and the distance are assumed to be $16$ K
and $D=11.5$ Mpc, respectively (Vidal-Madjar et al.\ 2000).
With these values, we obtain $f_\nu$ at 100 $\mu$m as
7 mJy. This flux level will be marginally observable with 
future space missions, such as ASTRO-F
(in the pointing mode; detailed information of this
mission is available in http://www.ir.isas.ac.jp/ASTRO-F/index-e.html.)
and {SIRTF} (Bicay, Werner 1998). The future infrared
facilities are summarized in Appendix of Takeuchi et
al.\ (1999).

The temperature of 16 K is a conservative value based 
on the assumption that dust is heated by the mean interstellar
UV radiation field (Vidal-Madjar et al.\ 2000). 
If dust exists near star-forming regions,
the temperature should be higher.
If the radiation field is 30-times larger than that estimated in
Vidal-Madjar et al.\ (2000), the dust temperature becomes
$T=23$ K. In this case, we obtain $f_\nu$ at 100 $\mu$m as
100 mJy and the detection becomes easier. Thus, the flux is strongly
sensitive to the dust temperature. The temperature should also be
determined from multi-band photometry by ASTRO-F.

\section{Summary}

We have examined the expected upper limit of the ionization degree
in the most metal-deficient galaxy, I Zw 18, 
while adopting the recent data
provided by Vidal-Madjar et al.\ (2000). Moreover, we present
a critical ionization degree above which the H$^-$-process
can be dominant over the formation process on the grain surfaces.
This critical ionization degree is comparable to the upper limit
of the ionization degree ($\sim 2\times 10^{-3}$).
We thus expect that the H$^{-}$-process works for 
the formation of H$_{2}$. At least, both the H$^{-}$-process
and the dust-process can work together. 
In order to obtain further constraint
on the formation process of H$_2$, we estimated the flux of
the NIR H$_2$ emission line, free-free radio emission,
and a FIR dust continuum.
The flux level of the H$_2$ line emission is 10 $\mu$Jy, which
is detectable by ASTRO-F.  
If there are many {H{\sc i}} clumps in the H {\sc i} envelope,
we may detect the free-free emission of the free electrons in 
the radio bands. 
Furthermore, the FIR continuum
emission is also detected by ASTRO-F with a 10 mJy level if the 
dust-to-gas
ratio is about 1/50 of the Galactic value.
If the FIR flux is below this level, the dust-to-gas ratio may be
smaller than 1/50 of the Galactic value and
the lower bound of the ionization degree for the dominant
H$^{-}$-process is lowered (i.e., the possibility that
H$^{-}$-process is the dominant process for the H$_2$ formation
becomes higher).

\bigskip

We wish to thank the referee, Prof. J.H.Black, for invaluable comments
that substantially improved the discussion of the paper.
We are grateful to S. Mineshige for continuous encouragement.
We also thank A.K. Inoue for useful discussions. 
One of us (HH) acknowledges the Research Fellowship of the Japan
Society for the Promotion of Science for Young Scientists. We
fully utilized the
NASA's Astrophysics Data System Abstract Service (ADS).

\section*{References}
\small

Bicay, M. D., \& Werner, M. W. 1998, in  ASP Conf.\ Ser. 148,
ORIGINS, ed.\ C. E. Woodward, J. M. Shull, \& H. A. Thronson Jr. 
(San Francisco: ASP), 290

Black, J. H., \& van Dishoeck, E. F. 1987, ApJ, 322, 412

Dickey, J. M., \& Lockman, F. J. 1990, ARA\&A, 28, 215 

Draine, B. T., \& Lee, H. M. 1984, ApJ, 285, 89

Dufour, R. J., Garnett, D. R., \& Shields, G. A. 1988, 
ApJ, 332, 752

Frail, D. A., Weisberg, J. M., Cordes, J. M., \& Mathers, C. 
1994, ApJ, 436, 144

Gondhalekar, P.M., Johansson, L.E.B., Brosch, N., Glass, I.S., \&
Brinks, E.  1998, A\&A, 335, 152

Herbst, E.  1995, Ann.Rev.Phys.Chem., 46, 2

Hirashita, H.  1999, ApJ, 522, 220

Hollenbach, D. J., \& Tielens, A. G. G.M. 1999, Rev.Mod.Phys.,
71, 173 

Jenkins, E. B., \& Peimbert, A. 1997, ApJ, 477, 265

Jura, M. 1974, ApJ, 191, 375

Kulkarni, S. R., \& Heiles, C. 1988, in Galactic and Extragalactic 
Radio Astronomy, ed. G. L. Verschuur \& K. I. Kellermann 
(New York: Springer), 95

Legrand, F., Kunth, D., Roy, J.-R., Mas-Hesse, J. M., \& Walsh, J. R.
    2000, A\&A, 355, 891

Lequeux, J., \& Viallefond, F. 1980, A\&A, 91, 269
 
Lonsdale Persson, C. J., \& Helou, G. 1987, ApJ, 314, 513

Mathis, J. S., Mezger, P. G., \& Panagia, N.  1983, A\&A, 128, 212

Matsuda, T., Sato, H., \& Takeda, H. 1969, Prog.Theo.Phys., 42, 219

Martin, C. L. 1996, ApJ, 465, 680

Martin, C. L. 1997, ApJ, 491, 561 

Nishi, R., Susa, H., Uehara, H., Yamada, M., \& Omukai, K. 1998,
Prog.Theo.Phys., 100, 881

Omukai, K. 2000, ApJ, 534, 809

Onaka, T. 2000, in ISO beyond the Peaks: The 2nd ISO Workshop on
Analytical Spectroscopy, ed.\ A. Salama, M. F. Kessler,
K. Leech \& B.Schultz, ESA-SP 456, 361

Reynolds, R.J., Hausen, N.R., Tufte, S.L., \& Haffner, L.M. 1998,
ApJ, 494, L99

Rybicki, G. B., \& Lightman, A. P. 1979, Radiative Processes in
    Astrophysics (New York: Wiley)

Sait\={o}, M., Kamaya, H., \& Tomita, A. 2000, PASJ, 52, 821

Shibai, H. 2000, Adv.\ Sp.\ Res., 25, 2273

Spitzer L. Jr 1978, Physical Processes in the Interstellar Medium
(New York: Wiley) ch7

Stasi\'nska, G., \& Schaerer, D. 1999, A\&A, 351, 72

Takahashi, J., Masuda, K, \& Nagaoka, M. 1999, MNRAS, 306, 22

Takeuchi, T. T., Hirashita, H., Ohta, K., Hattori, T. G., Ishii, T. T.,
\& Shibai, H. 1999, PASP, 111, 288

van Zee, L., Westpfahl, D., Haynes, M.P., \& Salzer, J.J. 1998, 
AJ, 115, 1000

Vidal-Madjar, A., Kunth, D., Lecavelier des Etangs, A.,
Lequeux, J., Andr$\rm \acute{e}$, M., Benjaffel, L., Ferlet, R.,
H$\rm \acute{e}$brard, G., et al. 2000, ApJ, 538, L77

Williams, D. A. 1993, in Dust and Chemistry in Astronomy,
ed. T.J.Millar and D.A.Williams (Philadelphia, IOP Publishing), 143 

\end{document}